\documentclass{emulateapj}
\usepackage{amssymb}
\usepackage{amsmath}
\usepackage{mathrsfs}
\usepackage{natbib}
\usepackage[colorlinks, linkcolor=blue, urlcolor=blue, citecolor=blue]{hyperref}
\usepackage{array}
\usepackage{float}
\usepackage{graphicx}
\usepackage{subfigure}
\usepackage{rotating}
\usepackage{epstopdf}
\usepackage{color}
\usepackage[all]{hypcap}
\usepackage[toc,page]{appendix}
\usepackage{array}
\usepackage{booktabs}

\RequirePackage{lineno}
\begin{document}

\title{Emission from a pulsar wind nebula: application to the persistent radio counterpart of FRB 121102}
\author{Yu-Han Yang\altaffilmark{1,2} and Zi-Gao Dai\altaffilmark{1,2}}

\begin{abstract}
The properties of fast radio bursts (FRBs) indicate that the physical origin of this type of astrophysical phenomenon is related to neutron stars. The first detected repeating source, FRB 121102, is associated with a persistent radio counterpart. In this paper, we propose that this radio counterpart could arise from a pulsar wind nebula powered by a magnetar without surrounding supernova ejecta. Its medium is a stratified structure produced by a progenitor wind. The model parameters are constrained by the spectrum of the counterpart emission, the size of the nebula, and the large but decreasing rotation measure (RM) of the repeating bursts. In addition, the observed dispersion measure is consistent with the assumption that all of the RM comes from the shocked medium.
\end{abstract}

\keywords{pulsars: general -- stars: magnetic fields, neutron -- radio
continuum: general}

\affil{\altaffilmark{1}School of Astronomy and Space Science, Nanjing University, Nanjing 210093, China; dzg@nju.edu.cn}
\affil{\altaffilmark{2}Key Laboratory of Modern Astronomy and Astrophysics (Nanjing University), Ministry of Education, China}


\section{Introduction}

\label{sec:Intro} Fast radio bursts (FRBs) are millisecond transients of GHz radio radiation \citep{lorim07, kean12, thor13, spit14, masui15, ravi15,ravi16, cham16,  petr16}. Their physical origin is still mysterious. Their large dispersion measures ($\text{DMs} \sim 100-2600$ pc cm$^{-3}$\footnote[3]{\url{http://frbcat.org/}}) suggest that they are at cosmological distances \citep{ravi19}. The DM of an FRB contains multiple possible contributions, e.g., the disk and halo of the Milky Way and the host galaxy \citep{opper12, dolag15, xu15, tend17, yao17}, the intervening intergalactic medium \citep{mcq14, aka16}, and their local environment \citep{conn16, lyu16, piro16, yang17, michi18}.
FRB 121102 is the first detected repeating event among the observed FRBs \citep{scho16, spit16, chat17, marc17, gajj18, zhang18}.
\cite{chat17}, \cite{marc17} and \cite{tend17} discovered both radio and optical counterparts associated with FRB 121102 and identified its host galaxy at a redshift of $z=0.193$.

Quite a few studies have focused on the explosion model of this kind of repetitive FRB. For models in which FRBs are powered by a young neutron star formed in core-collapse supernova (SN; \citealp{popo10, nich17, wax17}), FRB signals may pass through the supernova remnant (SNR), which provides a portion of the DM and the rotation measure (RM). Although the contribution of this fraction of the DM is small, it dominates the evolution of the total DM over years \citep{katz16, mura16, piro16, megz17, yang17, yzh17}. However, the DM evolution has not been detected yet. In addition, both a wind bubble produced by an interaction between an ultra-relativistic pulsar wind and the ejecta \citep{mura16} and a forward shock (FS) produced by an interaction between the fast outer layer of SN ejecta and the ambient medium \citep{megz17} can provide persistent radio radiation. In order to avoid the DM evolution, an ultra-stripped SN with the ejecta of mass $\lesssim 0.1M_{\odot}$ was proposed \citep{pirokul13, kashi17}, or even no ejecta around the pulsar \citep{dai17}. What's more, if the progenitor is a massive star, it should have produced a strong, magnetized wind \citep{ignace98, ud02}, leading to a decreasing density profile \citep{cheva82, cheva03, harvey10}, which is different from a constant-density interstellar medium \citep{piro18}. 

The observed data of the persistent radio counterpart have not been fitted satisfactorily and some details have remained unclear. Motivated by these arguments above, in this paper we investigate the DM and RM seen for FRB 121102 and their time evolution based on the scenario that a magnetar is surrounded by a stellar wind medium rather than SN ejecta. The pulsar wind nebula (PWN) is produced by an interaction between a pulsar wind and its magnetized stellar wind environment. We show that our PWN scenario can explain the persistent radio counterpart in reasonable ranges of the model parameters. This paper is organized as follows. In Section 2, we describe the dynamics of our model. We discuss the fitting results for observations of this source in Section 3, and conclude with a summary of our work in Section 4.


\section{The model of a PWN}

\label{sec:mod}

We propose a PWN model for the persistent radio source of FRB 121102, in which an ultra-relativistic wind from an ultra-strongly magnetized pulsar interacts with an existing stellar wind. In order to validate this model, we consider two scenarios. First, for a single star, a massive star that has produced a stellar wind collapses to a neutron star and a highly anisotropic outflow is ejected. In this case, no SN ejecta would be expected along our line of sight. Second, for a binary system, alternatively, the progenitor is a white dwarf (WD) with a companion star. The mass loss of this companion leads to a stellar wind environment, in which the binary system is immersed, and eventually accretion-induced collapse (AIC) of the WD generates an ultra-strongly magnetized pulsar without any SN ejecta \citep{nomo91, yu15}.

\subsection{Dynamics}

\begin{figure}[tbph]
\begin{center}
\includegraphics[width=0.85\textwidth,angle=0]{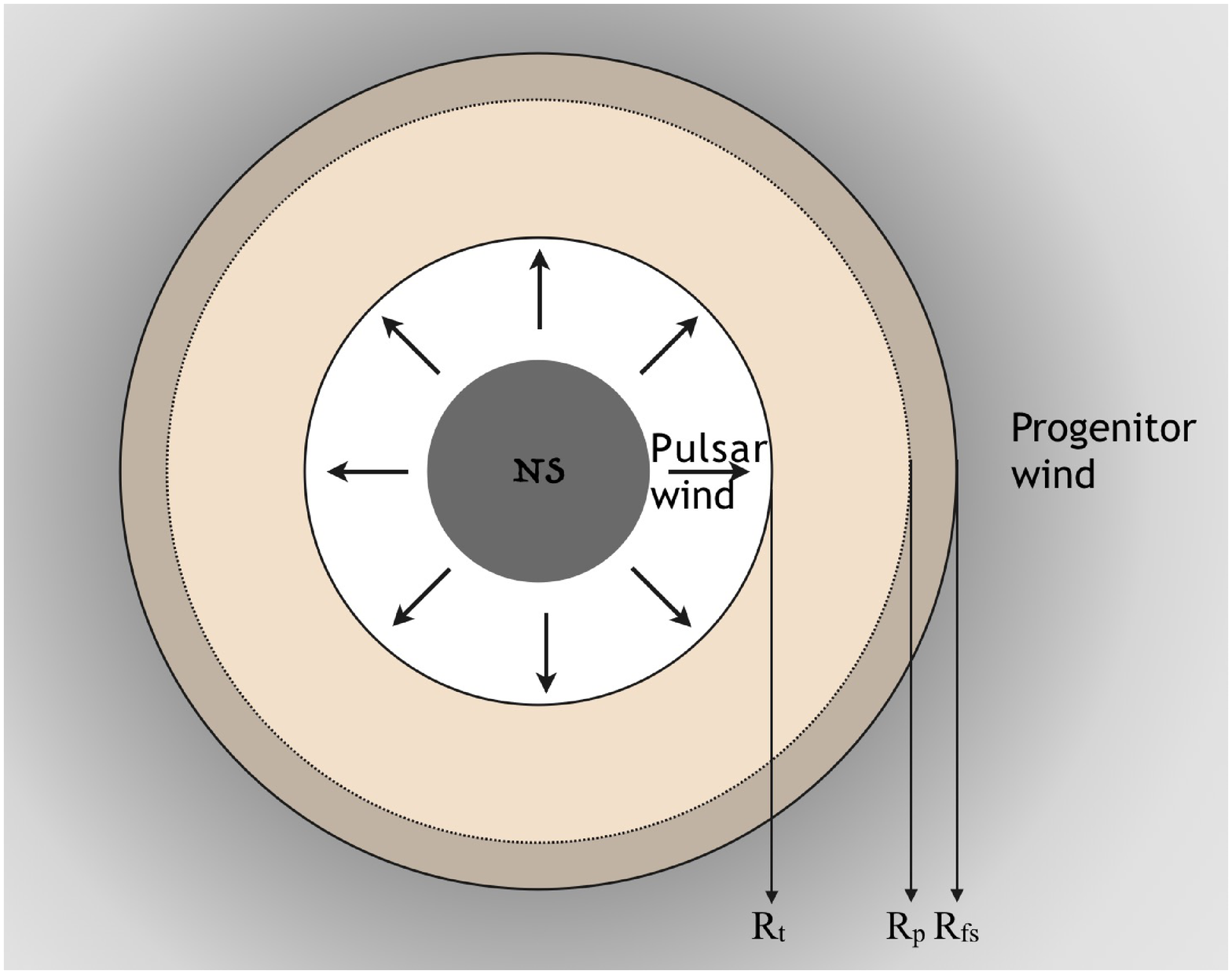}
\end{center}
\caption{The schematic picture of our model. A young pulsar at the center of the source, which generates an ultra-relativistic wind filled with electrons/positrons pairs. Assuming that there is no ejecta in our line of sight, the pulsar wind directly interacts with the circumstellar medium. This physical process forms two shocks: a reverse shock (RS) heating the cold wind, and a forward shock (FS) that goes into the surrounding medium. So the PWN consists of four regions, including the unshocked medium (Region 1), the shocked medium swept up by the FS (Region 2), the shocked wind swept up by the RS (Region 3), and the unshocked wind (Region 4). }
\label{fig:schema}
\end{figure}

A young pulsar generates an ultra-relativistic wind that is dominated by electrons/positrons pairs. Considering that there is no ejecta along our line of sight, the pulsar wind directly interacts with the surrounding environment.
Due to this wind sweeping up its surrounding medium, it results in two shocks: a reverse shock (RS) heating the cold wind, and an FS propagating into the surrounding medium. Therefore, this PWN consists of four regions: the unshocked medium (Region 1), the shocked medium swept by the FS (Region 2), the shocked wind swept by the RS (Region 3), and the unshocked wind (Region 4). The schematic figure of our model is presented in Figure \ref{fig:schema}.

The motion of the PWN satisfies the conservation of momentum \citep{raynolds84},
\begin{equation}  \label{eq:momc}
M_{\mathrm{sw}}\ddot{R}_{p}=4\pi R_p^{2}\left( P_{\mathrm{pwn}}-\rho _{0}\dot{R}_{p}^{2}\right),
\end{equation}
where ${R}_{p}$ and $P_{\text{pwn}}$ are the radius and pressure of the nebula, respectively, and $\rho _{0}$ is the density of the surrounding medium. We presume the scale of the shocked medium swept by the FS is much smaller than the radius of the PWN.

We assume that the density profile of the medium is a power-law function
\begin{equation}
	\rho _{0}=Ar^{-k}.
\end{equation}
If the ambient medium is a steady wind with a constant mass-loss rate from the progenitor, the index $k=2$ and $A=\frac{\dot{M}}{4\pi v_{w}}=5.0\times 10^{13}\dot{M}_{-5}v_{6}^{-1}$\,g\,cm$^{-1}$ , where $\dot{M}=10^{-5}\dot{M}_{-5}M_{\odot }$\,yr$^{-1}$ and $v_{w}=10^{6}v_6$\,cm\,s$^{-1}$ are the mass-loss rate and wind velocity of the progenitor, respectively \citep{piro18}. As a convention, we employ $Q=10^nQ_n$ in cgs units if there are no other explanations. The mass of shocked medium swept up by the FS is given by
\begin{equation}\label{eq:Msw}
	M_{\text{sw}}=\int_{0}^{R_{p}}4\pi r_{0}^{2}\rho _{0}dr=4\pi AR_{p}.
\end{equation}

The conservation of energy gives \citep{raynolds84, pirokul13}
\begin{equation}  \label{eq:enmc}
\frac{d}{dt}\left( 4\pi R_{p}^{4}P_{\text{pwn}}\right) =\left( L-\Lambda \right) R_{p},
\end{equation}
where $L$ is the luminosity of the pulsar, and $\Lambda $ is the radiative energy loss.
If we ignore the radiative energy loss, $\Lambda=0$, a self-similar solution of the radius of PWN can be obtained by Equations (\ref{eq:momc}) and (\ref{eq:enmc}),
\begin{equation}
	R_{p}=\left( \frac{L}{8\pi A}\right) ^{1/3}t=3.4\times 10^{-2}L_{40}^{1/3}A_{13}^{-1/3}t_{10\text{yr}}\text{ pc},
\end{equation}
where $t_{10\text{yr}}=t/10$ yr. The pressure in the nebula is
\begin{equation}
	P_{\text{pwn}}=\frac{A}{t^{2}}=1.0\times 10^{-4}A_{13}t_{10\text{yr}}^{-2}
\text{ dyn cm}^{-2}.
\end{equation}
The pulsar wind is terminated when the ram pressure of the wind is balanced by the PWN pressure \citep{gaen06}
\begin{equation}
	R_{t}\simeq \left( \frac{L}{4\pi cP_{\text{pwn}}}\right)^{1/2}=5.3 \times 10^{-3}L_{40}^{1/2}A_{13}^{-1/2}t_{10\text{yr}}\text{ pc},
\end{equation}
which satisfies the condition that $R_{t}\ll R_{p}$.

The total number of electrons in region 2 is written as
\begin{equation}
	N_{e,\mathrm{fs}}=M_{\mathrm{sw}}/m_{p}=\frac{4\pi AR_{p}}{m_{p}},
\end{equation}
where $m_{p}$ is the proton mass, and the subscript ``fs" indicates the corresponding quantities of the forward shocked medium (Region 2). In the following equations, the subscript ``rs" denotes the quantities of the reverse shocked wind (Region 3). The radius of Region 2 is
\begin{equation}\label{eq:nefsdelR}
	\Delta R_{p} = \frac{N_{e,\text{fs}}}{n_{e,\text{fs}}\times 4\pi R_{p}^{2}} = \frac{1}{n_{e,\text{fs}}} \frac{A}{m_{p}R_{p}} .
\end{equation}


\subsection{Emission}

We discuss synchrotron emission from Region 3. Considering that the particles in the nebula are relativistic, the energy density in Region 3 is
\begin{equation}
	U_{\text{rs}}=3P_{\text{pwn}}.
\end{equation}
The magnetic field of the PWN is expressed by
\begin{eqnarray}
	B_{\text{rs}} =\left( 8\pi \varepsilon _{B,\text{rs}}U_{\text{rs}}\right)^{1/2}=0.087A_{13}^{1/2}\varepsilon _{B,\text{rs}}^{1/2}t_{10\text{yr}}^{-1}
\text{ G,}
\end{eqnarray}
where $\varepsilon _{B,\text{rs}}$ is the ratio of the magnetic energy density to the total energy density behind the RS.

Electrons (and positrons) in the cold pulsar wind (Region 4) are accelerated to ultra-relativistic energies by the termination shock at $R_{t}$. We assume that the electron density profile is a power-law function initially, $dn_{e,\text{rs}}^{\prime }/d\gamma _{e}=K^{\prime }\gamma _{e}^{-p}$\,cm$^{-3}$ for $\gamma _{\min }\leq \gamma _{e}\leq \gamma _{\max }$. In this equation, $\gamma _{\min }$ and $\gamma _{\max }$ are the minimum and maximum Lorentz factor in Region 3, respectively,
\begin{eqnarray}
	\gamma_{\max }&\approx &\left( \frac{6\pi q_{e}}{\sigma _{T}B_{\text{rs}}}\right) ^{1/2}=4.0\times 10^{8}A_{13}^{-1/4}\varepsilon _{B,\text{rs}}^{-1/4}t_{10\text{yr}}^{1/2},\\
	\gamma _{\min } &=& \left( \frac{2-p}{p-1}\varepsilon _{e,\text{rs}}\gamma _{w}\gamma _{\max }^{p-2}\right) ^{1/\left( p-1\right)}+1 \notag  \\
	&\simeq & 51 \left(  A_{13} \varepsilon _{B,\text{rs}} \right) ^{ \frac{2-p}{4(p-1)}} (\varepsilon _{e,\text{rs}}\gamma _{w,5})^{ \frac{1}{p-1}}  t_{10\text{yr}}^{\frac{p-2}{2(p-1)} },
\end{eqnarray}
where $\gamma _w$ is the bulk Lorentz factor of the pulsar wind, $\varepsilon _{e,\text{rs}}=1-\varepsilon _{B,\text{rs}}$ is the electron energy density fraction of this Region \citep{dai01,huang06}, and the ``constant" coefficients in this equation and the following equations are taken for $p=1.53$. The cooling Lorentz factor in this Region is given by
\begin{equation}
	\gamma _{c}=\frac{6\pi m_{e}c}{\sigma _{T}B_{\text{rs}}^{2}t}=320A_{13}^{-1}\varepsilon _{B,\text{rs}}^{-1}t_{10\text{yr}},
\end{equation}
where $\sigma _{T}$ is the Thomson scattering cross section. The corresponding frequency is obtained by
\begin{eqnarray}\label{equ:nuc}
	\nu _{c} &=&\frac{\gamma _{c}^{2}}{1+z}\frac{q_{e}B_{\text{rs}}}{2\pi m_{e}c}\notag  \\
	&=&2.2\times 10^{10}A_{13}^{-3/2}\varepsilon _{B,\text{rs}}^{-3/2}t_{10 \text{yr}}\text{ Hz.}
\end{eqnarray}
Similarly, the corresponding frequency of the minimum Lorentz factor is written by
\begin{eqnarray}
	\nu _{\min }&=&\frac{\gamma _{\min }^{2}}{1+z}\frac{q_{e}B_{\text{rs}}}{2\pi m_{e}c}\notag \\
	&=&5.4\times10^{8}\left( A_{13}\varepsilon _{B,\text{rs}}\right)^{\frac{1}{2(p-1)} } \left(\varepsilon _{e,\text{rs}}\gamma _{w,5} \right)^{\frac{2}{p-1} } t_{10\text{yr}}^{\frac{1}{1-p} }\text{ Hz}. \notag \\
	&&
\end{eqnarray}

Due to the cooling effect, the distribution of the electrons behind the termination shock becomes \citep{sari98}
\begin{equation}
\frac{dn_{e,\text{rs}}}{d\gamma _{e}}=\left\{
\begin{array}{ccc}
K\gamma _{e}^{-p} & , & \gamma _{\min }\leq \gamma _{e}\leq \gamma _{c}\text{,} \\
K\gamma _{c}\gamma _{e}^{-\left( p+1\right) } & , & \gamma _{c}<\gamma _{e}\leq \gamma _{\max }\text{.}
\end{array}
\right.
\end{equation}
The total electron number density is found by
\begin{equation}
	n_{e,\text{rs}}=\frac{K}{\left( p-1\right) \gamma _{\min }^{p-1}}.
\end{equation}
Thus the election energy density is
\begin{eqnarray}
	U_{e,\text{rs}} &=&\varepsilon _{e,\text{rs}}U_{\text{rs}}=\int_{\gamma _{\min }}^{\gamma _{\max }}\frac{dn_{e,\text{rs}}}{d\gamma _{e}}\left( \gamma _{e}m_{e}c^{2}\right) d\gamma _{e}  \notag \\
	&\approx &\frac{K\gamma _{c}^{2-p}}{\left( 2-p\right) \left( p-1\right) }m_{e}c^{2}.
\end{eqnarray}
We then obtain
\begin{eqnarray}
	K &=&\frac{\left( 2-p\right) \left( p-1\right) \varepsilon _{e,\text{rs}}U_{ \text{rs}}}{m_{e}c^{2}\gamma _{c}^{2-p}}  \notag \\
	&=&6.1A_{13}^{3-p}\varepsilon _{B,\text{rs}}^{2-p}\varepsilon _{e,\text{rs} }t_{10\text{yr}}^{p-4}\text{ cm}^{-3}.
\end{eqnarray}
If $\nu _{\min }\leq \nu _{a}<\nu _{c}$, the synchrotron self-absorption frequency is written by
\begin{eqnarray}
	\nu _{a} &=&\frac{1}{1+z}\left( \frac{c_{2}q_{e}K_{\text{rs}}R_{p}}{B_{\text{rs}}}\right) ^{2/\left( p+4\right) }\frac{q_{e}B_{\text{rs}}}{2\pi m_{e}c} \notag \\
	&=&1.5\times 10^{9}\left( L_{40}^{\frac{2}{3}}A_{13}^{\frac{38-9p}{6}}\varepsilon _{B,\text{rs}}^{\frac{10-3p}{2}}\varepsilon _{e,\text{rs}}^{2}t_{10\text{yr}}^{p-8}\right) ^{\frac{1}{p+4}}\text{ Hz.}  \notag \\
	&&
\end{eqnarray}

The peak flux density at a luminosity distance of $D_{L}$ from the source is calculated by
\begin{eqnarray}
	F_{\nu ,\max } &=&\frac{\left( 1+z\right) N_{e,\text{rs}}m_{e}c^{2}\sigma _{T}}{4\pi D_{L}^{2}\times 3q_{e}}B_{\text{rs}}  \notag \\
	&=& 260L_{40}\left( A_{13} \varepsilon _{B,\text{rs}} \right) ^{\frac{8-3p}{4} }\gamma _{w,5} ^{-1} t_{10\text{yr}}^{\frac{p-2}{2} }\text{ }\mu \text{Jy,}
\end{eqnarray}
where
\begin{equation}
	N_{e,\text{rs}}=\frac{4\pi R_{p}^{3}n_{e,\text{rs}}}{3}.
\end{equation}

For the emission from Region 2 that is much fainter than that from Region 3, the synchrotron emission flux density at any frequency $\nu $ is given by
\begin{equation}
	F_{\nu }=\left\{
	\begin{array}{ll}
		F_{\nu ,\max }\left( \frac{\nu _{a}}{\nu _{\min }}\right) ^{-\frac{p-1}{2}}\left( \frac{\nu }{\nu _{a}}\right) ^{\frac{5}{2}}, & \nu <\nu _{a}\text{,} \\
		F_{\nu ,\max }\left( \frac{\nu }{\nu _{\min }}\right) ^{-\frac{p-1}{2}}, &  \nu _{a}<\nu <\nu _{c}\text{,} \\
		F_{\nu ,\max }\left( \frac{\nu _{c}}{\nu _{\min }}\right) ^{-\frac{p-1}{2}}\left( \frac{\nu }{\nu _{c}}\right) ^{-\frac{p}{2}}, & \nu \geq \nu _{c} \text{.}
	\end{array}
	\right.
\end{equation}
where the coefficient
\begin{eqnarray} \label{eq:Fmax1}
	F_{\nu ,\max }\left(\frac{\nu _{\min }}{\text{Hz} } \right)^{\left( p-1\right) /2} &=&5.3\times 10^{4}L_{40}\varepsilon _{e,\text{rs}}  \notag  \\
	&&\times \left( A_{13}^{\frac{3}{2}}\varepsilon _{B,\text{rs}}^{\frac{3}{2}}t_{10\text{yr}}^{-1} \right) ^{\frac{3-p}{2}}\text{ }\mu \text{Jy.}  \notag \\
	&&
\end{eqnarray}

Because Region 2 and 3 are separated by a contact discontinuity with a radius of $R_{p}$, the pressures should satisfy
\begin{equation}
	P_{\text{fs}}=P_{\text{rs}}.
\end{equation}
We infer that most of the electrons in Region 2 are in the non-relativistic regime for $\gamma _{p}=(1-v_{p}^{2}/c^{2})^{-1/2}\sim 1$, so we can ignore the synchrotron emission in Region 2, and the adiabatic index in Region 2 is 5/3, leading to a correlation of the energy densities,
\begin{equation}
	U_{\text{fs}}=\frac{1}{2}U_{\text{rs}}.
\end{equation}
Thus, the magnetic field in this Region is given by
\begin{eqnarray}
	B_{\text{fs}} &=&\left( 8\pi \varepsilon _{B,\text{fs}}U_{\text{fs}}\right)^{1/2}  \notag \\
	&=&620A_{13}^{1/2}\varepsilon _{B,\text{fs},-4}^{1/2}t_{10\text{yr}}^{-1} \text{ }\mu \text{G,}
\end{eqnarray}
where $\varepsilon _{B,\text{fs}}$ is the ratio of the magnetic energy density to the total energy density behind the FS.

\subsection{DM and RM}
Based on the electron number density of Region 3 and the dynamics of Region 2, we can deduce the DM and RM in the two regions.
The DM in Region 3 is calculated by
\begin{eqnarray}
	\text{DM}{\text{rs}} &=&\int_{0}^{R_{p}}n_{e,\text{rs}}dl=n_{e,\text{rs}}R_{p}  \notag \\
	&=&4.9\times 10^{-2} L_{40}^{\frac{1}{3}}A_{13}^{\frac{13}{6}-\frac{3p}{4}}\varepsilon _{B,\mathrm{rs}}^{\frac{3}{4}(2-p)}  \notag \\
	&&\times \gamma _{w,5} ^{-1} t_{10\text{yr}}^{-2+\frac{p}{2}}\text{ pc cm}^{-3},  \label{eq:DMrs}
\end{eqnarray}
which potentially changes with time. Taking the time derivative of Equation (\ref{eq:DMrs}), we find
\begin{eqnarray}\label{eq:dDMrs}
	\frac{d\text{DM}_{\text{rs}}}{dt} &=&-6.1\times 10^{-3}L_{40}^{\frac{1}{3}}A_{13}^{\frac{13}{6}-\frac{3p}{4}}\varepsilon _{B,\mathrm{rs}}^{\frac{3}{4}(2-p)}  \notag \\
	&&\times \gamma _{w,5} ^{-1} t_{10\text{yr}}^{-3+\frac{p}{2}}\text{ pc cm}^{-3} \text{ yr}^{-1} .
\end{eqnarray}

From Equation (\ref{eq:nefsdelR}), the DM in Region 2 is given by
\begin{eqnarray}\label{eq:DMfs}
	\text{DM}_{\text{fs}} &=&n_{e,\text{fs}}\Delta R_{p}  \notag \\
	&=&18L_{40}^{-1/3}A_{13}^{4/3}t_{10\text{yr}}^{-1}\text{ pc cm}^{-3}.
\end{eqnarray}
Taking the time derivative of Equation (\ref{eq:DMfs}), we find
\begin{equation}
	\frac{d\text{DM}_{\text{fs}}}{dt}=-1.8L_{40}^{-1/3}A_{13}^{4/3}t_{10\text{yr}}^{-2}\text{ pc cm}^{-3}\text{ yr}^{-1}.
\end{equation}

Since the electrons and the positrons in the shocked nebula produce the opposite RM, we just consider the RM of Region 2, which is
\begin{eqnarray}\label{eq:RM}
	\mathrm{RM}_{\text{fs}} &=&\frac{1}{\left( 1+z\right) ^{2}}\frac{q_{e}^{3}}{2\pi m_{e}^{2}c^{4}}\int_{0}^{R_{p}}n_{e, \text{fs}}B_{\text{fs}, \parallel }dl  \notag \\
	&\approx &6300L_{40}^{-1/3}A_{13}^{11/6}\varepsilon _{B,\mathrm{fs,-4}}^{1/2}t_{10\text{yr}}^{-2}\text{ }\mathrm{{rad\ cm}^{-2}.}
\end{eqnarray}
Both the DM and the RM decrease with time.

\section{Fitting Results}

\begin{figure}[tbph]
\begin{center}
\includegraphics[width=0.85\textwidth,angle=0]{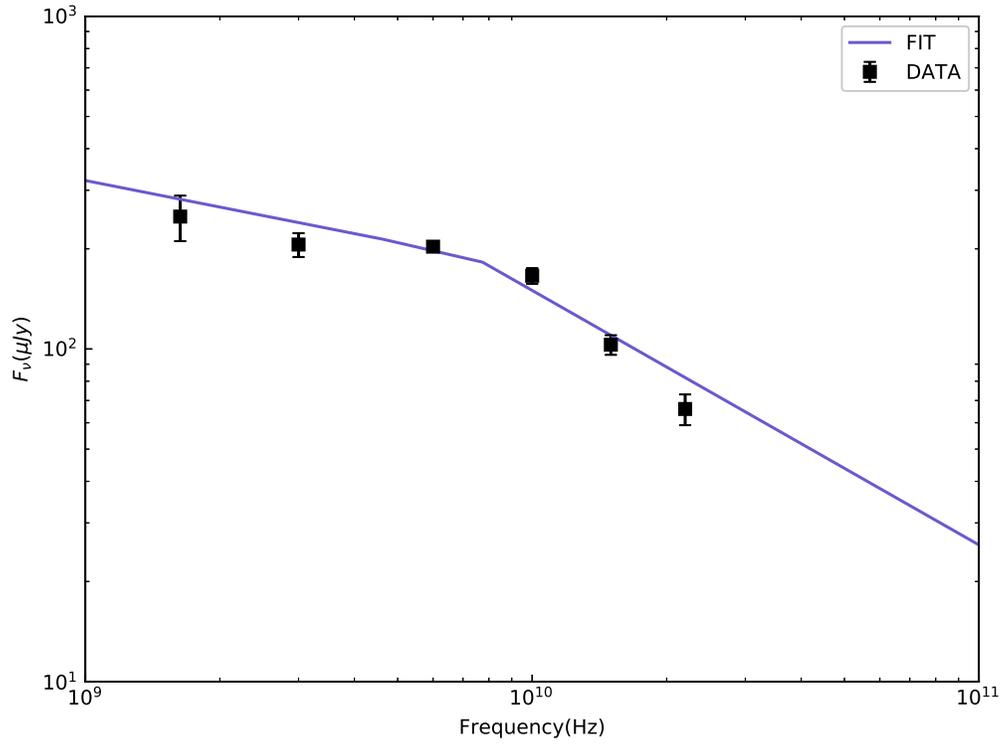}
\end{center}
\caption{The spectrum of the persistent radio counterpart of FRB 121102 with the parameters $F_{\nu ,\max }\left(\nu _{\min }/\text{Hz}  \right)^{\left( p-1\right) /2}=7.8\times 10^{4}$ $\mu$Jy, $p=1.53$ and $\nu _{c}=7.4\times 10^{9}$ Hz which are
the best value from the MCMC fitting. The black square data are observed by the Karl G. Jansky Very Large Array \citep{chat17}.
}
\label{fig:spectrum}
\end{figure}

In this section, we fit the observed results according to the model in Section 2, including the spectrum of the persistent radio counterpart and the RM observations accompanied by FRB 121102. We can fit the spectrum of the persistent radio source by using three parameters through the Markov chain Monte Carlo (MCMC) method, $F_{\nu, \mathrm{\max }}\left(\nu _{\min }/\text{Hz}  \right)^{(p-1)/2}$, $p$ and $\nu _c$. We obtain the best-fitting parameters: $F_{\nu ,\max }\left(\nu _{\min }/\text{Hz}  \right)^{\left( p-1\right) /2}=7.8\times 10^{4}$ $\mu$Jy, $p=1.53$ and $\nu _{c}=7.4\times 10^{9}{\rm \ Hz}$, and the corresponding spectrum is shown in Figure \ref{fig:spectrum}. By diffusive shock acceleration, the power-law index of relativistic electrons (whose adiabatic index is $\gamma = 4/3$) is $(r+2)/(r-1)=(3\gamma -1)/2=1.5$, where the shock compression ratio $r=(\gamma +1)/(\gamma -1)$ \citep{caprioli11}. This kind of hard electron spectrum has been applied to the radio emission component of many PWNe \citep[e.g.,][]{zhang08,martin16} and some gammay-ray burst (GRB) afterglows (e.g. GRB 000301c \citep{panaitescu01} and GRB 010222 \citep{dai01}).

Furthermore, we obtain the evolution time from Equation (\ref{equ:nuc})
\begin{equation}\label{equ:t0}
	t_{10\text{yr},0}=0.34A_{13}^{3/2}\varepsilon _{B\mathrm{,rs}}^{3/2},
\end{equation}
corresponding to MJD$=57637$ ($\sim 4$ yr later after the first pulse of FRB 121102 was detected). Introducing the evolution time when we observe the spectrum into Equation (\ref{eq:Fmax1}), the luminosity of the pulsar can be found by
\begin{equation}\label{eq:L}
	L_{40}=0.67\varepsilon _{e\mathrm{,rs}}^{-1}.
\end{equation}

Introducing the two conditions above, we get some constrains on these parameters. From the radio observation in 5\,GHz at MJD=57653, we know that $R_{p}\leq 0.7$ pc \citep{marc17},
\begin{equation}\label{eq:tup}
	A_{13}\leq 37\varepsilon _{B,\mathrm{rs}}^{-9/7}\varepsilon _{e,\mathrm{rs}}^{2/7}.
\end{equation}
From $\nu _{a}\leq 1.63$ GHz, we find
\begin{equation}\label{eq:tlow}
	A_{13}\geq 3.1\varepsilon _{B,\mathrm{rs}}^{-21/17}\varepsilon _{e,\mathrm{rs}}^{4/17}.
\end{equation}
By substituting Equation (\ref{eq:L}) into Equation (\ref{eq:RM}), the rotation measure is derived as
\begin{equation}\label{eq:RM1}
	\mathrm{RM}_{\mathrm{fs}}=7200A_{13}^{11/6}\varepsilon _{e,\mathrm{rs}}^{1/3}\varepsilon _{B,\mathrm{fs,-4}}^{1/2}t_{10\text{yr}}^{-2}\text{ }\mathrm{{rad\ cm}^{-2}.}
\end{equation}

\begin{figure}[tbph]
\begin{center}
\includegraphics[width=0.85\textwidth,angle=0]{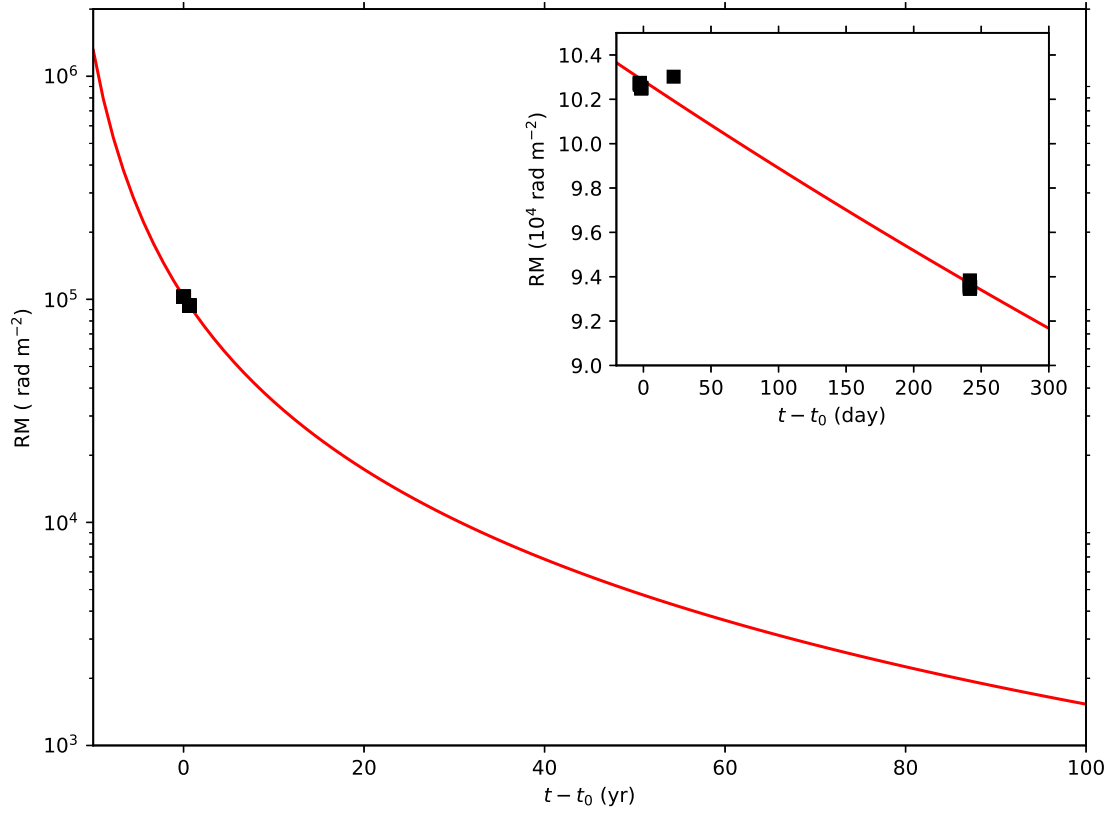}
\end{center}
\caption{Evolution of RM with time. The red line is corresponding to our model. The square data are the RM observed by Arecibo and the Robert C. Byrd Green Bank Telescope \citep{gajj18, michi18}.}
\label{fig:RM}
\end{figure}

The two values of the observed RM are RM$_{1}=1.0\times 10^{5}$ rad m$^{-2}$ (MJD=57750) and RM$_{2}=9.4\times 10^{4}$ rad m$^{-2}$ (MJD=57991; \citealp{gajj18, michi18}).
The corresponding evolution time of the first observed RM approximates to $t_0$.
\begin{table}
\centering
\caption{Physical Magnitudes}
\label{table:1}
\begin{center}
\setlength{\tabcolsep}{5mm}{
\begin{tabular}{lc}
\toprule
Quantities &  Reasonable values\\
\hline
\specialrule{0em}{3pt}{3pt}
$t_0$ (yr)&  14\\
$\varepsilon _{B,\text{rs}}$ & $0.55-0.99986$\\
$\varepsilon _{B,\text{fs}}$ & $5.1\times 10^{-4}-0.96$\\
$L$ (erg s$^{-1}$)& $1.5\times 10^{40}-4.7\times 10^{43}$\\
$A$ (g cm$^{-1}$)& $2.5\times 10^{13}-4.5\times 10^{13}$\footnote{The range of physical quantities marked correspond to the opposite range of $\varepsilon _{B,\text{rs}}$.\label{foot}}\\
\specialrule{0em}{3pt}{3pt}
\hline
\specialrule{0em}{3pt}{3pt}
DM$_{\text{rs}}$ ($\text{ pc cm}^{-3}$) & $0.14-1.4$\\
$\left |\frac{d\text{DM}_{\text{rs}}}{dt}\right |$ ($\text{ pc cm}^{-3} \text{ yr}^{-1}$)& $0.013-0.12$\\
\specialrule{0em}{3pt}{3pt}
DM$_{\text{fs}}$ ($\text{ pc cm}^{-3}$) & $2.6-85^{\text{\ref{foot}}}$\\
$\left|\frac{d\text{DM}_{\text{fs}}}{dt}\right|$ ($\text{ pc cm}^{-3} \text{ yr}^{-1}$)& $0.19-6.1^{\text{\ref{foot}}}$\\
\specialrule{0em}{3pt}{3pt}
\hline
\end{tabular}\\}
\end{center}
\leftline{\bf{Note.}}
\end{table}
Introducing these observed data into Equation (\ref{eq:RM1}), $t_{10\text{yr},0}=1.4$ can be obtained, which is consistent with $t_0 >4$ yr. Substituting this condition into Equation (\ref{equ:t0}), we can get
\begin{equation}\label{eq:A}
	A_{13}=2.5/\varepsilon _{B,\mathrm{rs}}.
\end{equation}
By substituting $t_{10\text{yr},0}$ and Equation (\ref{eq:A}) into Equation (\ref{eq:RM1}), we can obtain $\varepsilon _{B,\mathrm{fs},-4}=26\varepsilon _{B,\mathrm{rs}}^{11/3}\varepsilon _{e,\mathrm{rs}}^{-2/3}$. Since $\varepsilon _{B,\mathrm{fs},-4}<1$ and by introducing Equation (\ref{eq:A}) into the limit of $A_{13}$, we can infer the limit that $0.55\leq \varepsilon _{B,\mathrm{rs}}\leq (1-1.4\times 10^{-4})$. Then the ranges of all parameters could be obtained, which are presented in Table \ref{table:1}. For a neutron star, if the progenitor or the companion of its progenitor is a massive star, which has typical mass-loss rate and wind velocity, we can obtain the range of the wind mass loading parameter ($A$) in the Table. DM$_{\text{rs}}$ and $|d\text{DM}_\text{rs}/dt|$ are calculated with $\gamma _w =10^5$. If we choose a smaller $\gamma _w$, the contributions of Region 2 to the DM and its evolution would be more important. As $\varepsilon _{B,\mathrm{rs}} $ increases, DM$_\text{rs}$ and $|d\text{DM}_\text{rs}/dt|$ increase, but DM$_{\text{fs}}$ and $|d\text{DM}_\text{fs}/dt|$ decrease. We plot the RM evolution in Figure \ref{fig:RM}. It is seen from this figure that our model can explain the observed RM very well.

\begin{figure}[tbph]
\begin{center}
\includegraphics[width=0.85\textwidth,angle=0]{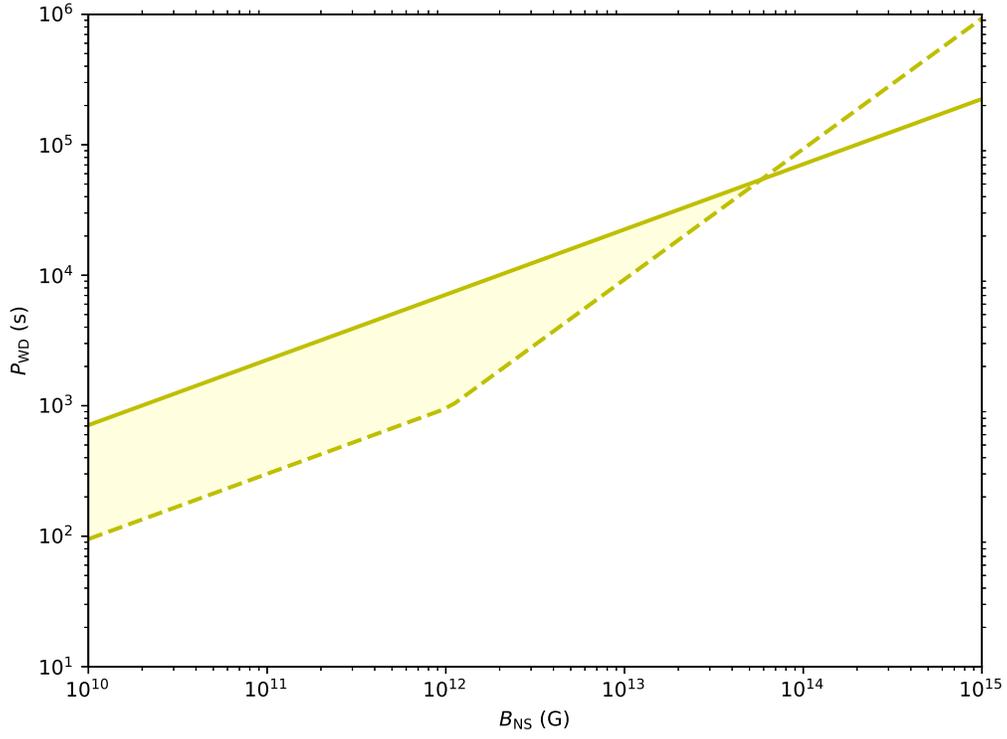}
\end{center}
\caption{Constraints on the period of a white dwarf progenitor $P_{\text{WD}}$ and the magnetic field of a neutron star $B_{\text{NS}}$. The neutron star is assumed to arise from accretion-induced collapse of the WD progenitor. The solid line results from the lower limit of the spin-down luminosity of the neutron star ($L_{\rm sd}$), and the dashed line is obtained from the upper limit of $L_{\rm sd}$ and the minimum value of $t_{\rm sd}$.}
\label{fig:Plim}
\end{figure}

By introducing the parameters in Table \ref{table:1} into Equation (\ref{eq:Msw}), we find that the shocked medium mass is $M_{\rm sw} \simeq 0.03 $ $M_{\odot }$. The scenario of AIC suggested in the first paragraph of Section 2 is valid if the SN ejecta mass is smaller than this value of $M_{\rm sw}$. In numerical simulations on AIC, the ejecta mass ($M_{\rm ej}$) was found to be in the range of a few times $10^{-3}$ $M_{\odot }$ to $\sim 0.1$ $M_{\odot }$ \citep{Dessart06,Darbha10,Ruiter19}. In order to give a low ejecta mass ($M_{\rm ej}<M_{\rm sw}$) that guarantees our model to be self-consistent, the WD progenitor must have a low initial poloidal magnetic field (e.g., $\lesssim10^{11}$ G; \citealp{Dessart07}) and be uniformly rotating slowly \citep{Abdikamalov10}.
Besides, in order to satisfy the assumption that the spin-down luminosity of the pulsar ($L=L_{\text{sd,}0}$) is constant, the evolution time $t_0$ should be less than the spin-down timescale $t_{\text{sd}}$. The characteristic spin-down luminosity and timescale due to magnetic dipole radiation are expressed by \citep{dai17}
\begin{eqnarray}
 L_{\text{sd,}0}&=&3.8\times 10^{43}\text{ }B_{\text{NS},12}^{2}P_{\text{NS},-3}^{-4}R_{\text{NS},6}^{6} \text{ erg s}^{-1},\\
 t_{\text{sd}}&=&16\text{ }I_{\text{NS},45}B_{\text{NS},12}^{-2}P_{\text{NS},-3}^{2}R_{\text{NS},6}^{-6} \text{ yr},
\end{eqnarray}
where $B_{\text{NS}}$, $P_{\text{NS}}$, $R_{\text{NS}}$ and $I_{\text{NS}}$ are the polar magnetic dipole field strength on the surface, the initial rotation period, the radius, and the moment of inertia of the neutron star.
The constraint on $t_0$ mentioned above and the range of the luminosity in Table \ref{table:1} can be used to obtain the limits on the initial rotation period of the neutron star,
\begin{eqnarray}
\max &[0.93I_{\text{NS},45}^{1/2}B_{\text{NS},12}R_{\text{NS},6}^{3},\,0.95B_{\text{NS},12}^{1/2}R_{\text{NS},6}^{3/2}]\text{ ms} \nonumber\\
 & <P_{\text{NS}}<7.1B_{\text{NS},12}^{1/2}R_{\text{NS},6}^{3/2}\text{ ms.}
\end{eqnarray}
According to the angular momentum conservation during the collapse of the WD (the effect of the low-mass ejecta is ignored), we find $\frac{2}{5}M_{\text{WD}}R_{\text{WD}}^{2}(2\pi /P _{\text{WD}})\sim \frac{2}{5}M_{\text{NS}}R_{\text{NS}}^{2}(2\pi /P_{\text{NS}})$, and thus the rotation period of the WD progenitor $P_{\text{WD}}$ is constrained in Figure \ref{fig:Plim}.
For the typical parameters ($R_{\text{WD}}\sim 10^9$ cm and $R_{\text{NS}}\sim 10^6$ cm), $P_{\text{WD}}$ is in the range of a few times $10^2-10^4$\,s.

\section{Summary}

\label{sec:dis}
In this paper, we have proposed a model for the persistent radio source associated with FRB 121102, in which a rapidly-rotating strongly magnetized pulsar's wind drives a non-relativistic PWN in an ambient progenitor wind. The PWN is produced by an interaction between the pulsar wind consisting of ultra-relativistic electron/positron pairs and the stellar wind. The spectrum of the persistent radio counterpart and the RM of FRB 121102 have been fitted in our model. The parameters in the model are constrained effectively. Furthermore, we obtained the evolution of the DM and RM.
Our model can still give a reasonable explanation for the observations.

This model has been constructed in two ways, which are scenarios of highly anisotropic ejecta and AIC of a WD. In the second case, the WD progenitor needs a low initial poloidal magnetic field, a small differential rotation, and a rotation period in the range of a few times $10^2 - 10^4\,$s.

Our model with an electron--positron component pulsar wind is more general than the model that assumes that pulsar wind contains a large number of ions \citep{margalit18}. In addition, our model is very simple. Compared with those models containing ejecta \citep[e.g.][]{margalit18, piro18}, this simple model does not provide very large DM evolution and free-free absorption. In addition, we have obtained a set of parameter values that can fully fit the observed data by restricting the model from the observations.

\acknowledgments

We thank Liang-Duan Liu for your helpful suggestions and Gaoyuan Zhang, Xiaotian Xu, Lei Sun, and Xiao Zhang for discussions. This work is supported by the National Key Research and Development Program of China (grant No. 2017YFA0402600) and the National Natural Science Foundation of China (grant No. 11573014 and 11833003).

\clearpage

\end{document}